\documentclass[english,aps,manuscript]{revtex4}
\usepackage[latin1]{inputenc}
\usepackage{graphicx}

\makeatletter

\newcommand{\bee}{\begin{equation}}
\newcommand{\ee}{\end{equation}}
\newcommand{\beea}{\begin{eqnarray}}
\newcommand{\eea}{\end{eqnarray}}

\usepackage{babel}
\makeatother
\begin{document}
\begin{center}\textbf{\Large Effective Potentials for Light Moduli}
\vspace{0.3cm}\end{center}

\begin{center}{\large S. P. de Alwis$^{\dagger}$ }\end{center}{\large \par}

\begin{center}Physics Department, University of Colorado, \\
 Boulder, CO 80309 USA\end{center}

\begin{center}\vspace{0.3cm}\end{center}

\begin{center}\textbf{Abstract}\end{center}

\begin{center}We examine recent work on compactifications of string
theory with fluxes, where effective potentials for light moduli have
been derived after integrating out moduli that are assumed to be heavy
at the classical level, and then adding non-perturbative (NP) corrections
to the superpotential. We find that this two stage procedure is not
valid and that the correct potential has additional terms. Althought
this does not affect the conclusion of Kachru et al (KKLT) that the
Kaehler moduli may be stabilized by NP effects, it can affect the
detailed physics. In particular it is possible to get metastable dS
minima without adding uplifting terms.\end{center}

\vspace{0.3cm}


\vfill

$^{\dagger}$ {\small e-mail: dealwis@pizero.colorado.edu}{\small \par}

\eject

\section{Introduction}

There has been much progress recently in understanding the mechanisms
by which the compactification moduli and the dilaton of string theory
are stabilized%
\footnote{For recent reviews with references to the original literature see
\cite{Balasubramanian:2004wx} \cite{Silverstein:2004id}.%
}. In particular Giddings et al. \cite{Giddings:2001yu} discussed
type IIB compactification on a Calabi-Yau orientifold $X$, and showed
that turning on fluxes of NSNS and RR three and five form fields can
generate a potential for the complex structure moduli of $X$ and
the dilaton-axion field, which as one would expect is of the supergravity
form. However the Kaehler moduli (which includes the overall scale
modulus usually denoted by $T$ in the string-phenomenology literature)
cannot be fixed by the fluxes. A suggestion for fixing these was subsequently
made by Kachru et al. \cite{Kachru:2003aw} (KKLT).

The KKLT proposal was to argue that at least for certain choices of
fluxes the dilaton-axion ($S$) and the complex structure moduli ($z^{i}$)
would have masses that are close to the string scale and could be
integrated out classically to get a theory for the light Kaehler moduli.
But at the classical minimum, the potential being of the no-scale
type, is zero and does not fix $T$ (for simplicity we will just consider
the case of one Kaehler modulus). In order to get a potential for
$T$ these authors proposed that a contribution coming from certain
non-perturbative effects be included. Concretely the argument is that
the fluxes which fix $S$ and $z^{i}$ give a constant $W_{0}$ in
the superpotential, to which the exponential contribution coming from
the non-perturbative (NP) effects should be added, resulting in a
total superpotential \[
W=W_{0}+Ce^{-aT},\]
 for the theory of the light modulus $T$. Here the pre-factor of
the NP term is taken to be a constant since we are ignoring perturbative
corrections as in KKLT. The Kaehler potential of the theory is taken
to be its classical value \[
K=-3\ln(T+\bar{T}).\]
 With this prescription it is easy to see \cite{Kachru:2003aw} that
the modulus $T$ is fixed at a supersymmetric Anti-deSitter (AdS)
point. KKLT go on to lift this minimum by adding a term coming from
$\bar{D}$ branes. 

In this note we will examine the consistency of the assumptions that
lead to the KKLT theory for the light modulus. These include the expectation
that the low energy supergravity action is a good starting point for
finding clssical string vacua and also that one can find flux configurations
such that the compolex stucture moduli and the dilaton are heavy.
As KKLT observed, in order for the procedure to make sense, the minimum
of the potential should be at a large value of $T$, so that the size
of $X$ is large on the string scale justifying the ten-dimensional
SUGRA starting point, and the superpotential itself is valid only
in the region $aT_{R}>>1$, so that the NP term can be regarded as
a small correction to the classical theory. KKLT argued that although
generically the fluxes would give a value of $W_{0}$ that is of order
one (thus violating the first requirement) there would be (at least
for CY manifolds $X$with large $h_{21}$) flux configurations which
would give small values of this constant.

Now of course there are obvious corrections to this theory coming
from perturbative effects which, even though they leave the superpotential
unchanged, will affect the Kaehler potential. There is also a non-perturbative
correction to the Kaehler potential (see for example \cite{Kaplunovsky:1994fg}\cite{Burgess:1995aa}).
Thus one would expect a corrected Kaehler potential of the form \begin{equation}
K=-3\ln(T+\bar{T}+f+ke^{-a(T+\bar{T})})\label{kahlernp}\end{equation}
 where $f$ is a constant and in principle the coefficient $k$ could
be of the same order as the prefactor $C$ in the superpotential.
We will ignore these corrections in most of this paper and will touch
on their effects at the end of our discussion. What we are going to
investigate is just the procedure of first ignoring the non-perturbative
term in the superpotential in order to integrate out $S,\, z^{i}$
to get a constant superpotential, and then including the non-perturbative
term. We will find that if the non-perturbative term is included from
the beginning, there are terms which are necessarily controlled by
the same coefficient as the terms which are included by KKLT and therefore
cannot be set to zero. Related obeservations have been made by Choi
et al. \cite{Choi:2004sx} %
\footnote{See also \cite{Lust:2005dy}. Related issues in the heterotic context
have been discussed in \cite{Curio:2005ew}.%
} but we will find that there needs to be some modifications of their
arguments also. In the course of this investigation we came across
some issues in the procedure of integrating out heavy fields in supersymmetric
theories that are of general interest, but we will reserve that discussion
to a separate publication \cite{deAlwis:2005tg}.

The two stage calculation of KKLT appears to lead only to a critical
point that is AdS supersymmetric. To get a dS minimum (and broken
supersymmetry) KKLT add an uplifting terms namely a contribution from
a Dbar brane. An alternate suggestion is to find some sector that
gives a D-term \cite{Burgess:2003ic}. However it is easy to see that
such a term will not lift an AdS supersymmetric minimum. This is because
of the relation\[
2\Re f^{ab}D_{b}=\frac{ik^{ai}D_{i}W}{W}\]

(with $k^{ai}$ a generator of a Killing symmetry of the Kahler metric
and $f$ the gauge coupling function) between the D and F terms that
is valid at generic points where the superpotential is non-zero %
\footnote{This relation can be found for example in \cite{Gates:1983nr} eqns.
(8.7.7b) (8.7.8). It has been recently rediscovered in the current
context in \cite{Choi:2005ge}. The right hand side of this equation
can also be rewritten as $ik^{ai}\partial_{i}K+\xi{\rm tr}T^{a}$
(with $\xi$a FI parameter) and would give an independent condition
if $W=0$.%
}. So a critical point where the F-term is zero with $W\ne0$ giving
an AdS minimum as in KKLT will not be lifted by adding a D term. This
also means that a Dbar term as in KKLT if it is to lift the AdS minimum
would have to be an explicit breaking term from the point of view
of four dimensional supergravity. Of course one can lift a non supersymmetric
$(D_{i}W\ne0)$ AdS critical point (such an example can be found in
\cite{Brustein:2004xn}) by a D-term. 

One of the outcomes of the current investigation is that if one integrates
out the heavy moduli in one stage then one has extra terms in the
potential (compared to the two stage procedure). These terms enable
one to find examples where the F-term potential by itself has positive
local minima thus obviating the need for uplifting terms.%
\footnote{It is possible that $\alpha'$corrections also achieve the same end.
See for instance \cite{Balasubramanian:2004uy}.%
} In view of this it would be interesting to revisit other issues such
as the question of getting a viable cosmology in the context of such
models.

\section{Model with S and T}

Consider first a compactification with fluxes on a rigid CY manifold
$X$, i.e. one with $h_{21}=0$. 

The classical Kaehler potential is\begin{equation}
K=-\ln(S+\bar{S)}-3\ln(T+\bar{T}),\label{KahlerST}\end{equation}
 and with the non-perturbative contribution included we have for the
superpotential,\begin{equation}
W=A+SB+Ce^{-aT},\label{WST}\end{equation}
 where $A,B$ are determined by the fluxes and $C$ is an $O(1)$
prefactor which may be determined by an instanton calculation. Now
let us first solve for $S$ in terms of $T$ as in KKLT by requiring
that the Kaehler derivative with respect to S of this superpotential
is zero %
\footnote{There are some issues involved in integrating out heavy fields in
supersymmetric theories that have not been discussed in the literature.
In particular it turns out that even in global supersymmetric theories
the condition $\partial_{H}W=0$ which is imposed in order to integrate
out a heavy field $H$ is valid only if we also restrict the light
field space to range over values that are less than the mass of the
heavy field. A similar restriction holds in supergravity. These issues
are discussed in a recent paper by the author \cite{deAlwis:2005tg}.
In particular it is shown there that solving the Kaehler derivative
equated to zero for the heavy field is an acceptable method of computing
the bosonic effective potential for the light fields even though to
get the complete action for the light fields addtional terms involving
the fermions need to be kept.%
}. This gives,\begin{equation}
D_{S}W=B-\frac{A+SB+Ce^{-aT}}{S+\bar{S}}=0,\label{Sderivative}\end{equation}

implying \begin{equation}
\bar{S}=(A+Ce^{-aT})/B.\label{SbarT}\end{equation}

Clearly if we substitute this back into (\ref{WST}) we get an expression
which contains both $T$and $\bar{T}$ i.e. it is not holomorphic.
As far as the scalar potential goes this is not a problem - the coupling
of the chiral scalars to supergravity is actually determined by one
Kaehler invariant real function (see for example \cite{Wess:1992cp}
or \cite{Gates:1983nr})\begin{equation}
G=K(\Phi,\bar{\Phi})+\ln W(\Phi)+\ln\bar{W}(\bar{\Phi}).\label{G}\end{equation}
 After solving for $S$ this becomes\begin{eqnarray}
G & = & -\ln(\frac{(A+Ce^{-aT})}{B}+\frac{(\bar{A}+\bar{C}e^{-a\bar{T}})}{\bar{B}})-3\ln(T+\bar{T)}\nonumber \\
 &  & +\ln(A+B\frac{(\bar{A}+\bar{C}e^{-a\bar{T}})}{\bar{B}}+Ce^{-aT})+c.c.).\label{Gcorrect}\end{eqnarray}

This may in effect be regarded as the new Kaehler potential with the
superpotential being taken to be unity. The potential for $T$ may
now be computed from the standard formula\begin{equation}
V=e^{G}(G_{i}G_{\bar{j}}G^{i\bar{j}}-3),\label{potential}\end{equation}
 where $G_{i}=\partial G/\partial\Phi^{i}$ and $G_{i\bar{j}}=\partial_{i}\partial_{\bar{j}}G$
is the Kaehler metric. 

On the other hand if we had followed the prescription of KKLT we would
have solved for $S$ in the absence of the non-perturbative term to
get $\bar{S}=A/B$, a constant superpotential $W=A+B\bar{A}/\bar{B}\equiv W_{0}$
and apart from an irrelevant constant the Kaehler potential is $K=-3\ln(T+\bar{T})$.
Now the non-perturbative term is added to $W$ to get a Kaehler invariant
function,\begin{equation}
G=-3\ln(T+\bar{T})+(\ln(A+B\frac{\bar{A}}{\bar{B}}+Ce^{-aT})+c.c.)\label{GKKLT}\end{equation}

The problem is that in this two stage process one is ignoring non-perturbative
terms in (\ref{Gcorrect}) that are in fact controlled by the same
constant $C$ as the terms that are being kept. \emph{There is no
approximation in which one can keep the latter and ignore the former.}
In other words the procedure of first integrating out $S$ and then
adding the non-perturbative term to $W$ cannot be justified. It should
be noted also that this correction is of the same order as the term
kept by KKLT independently of the condition $W_{0}<<1$ required by
KKLT.

\section{Models with complex structure moduli}

The model studied in the previous section however does not give a
viable theory in any case. Choi et al \cite{Choi:2004sx} have analyzed
the stability of this model (without first integrating out the dilaton-axion
$S$). They find that the supersymmetric extremum $D_{S}W=D_{T}W=0$
is in fact a saddle point. Although this does not make the supersymmetric
point unstable (since a saddle point or even a maximum can be a stable
AdS solution) it becomes problematic when one adds a {}``lifting
potential'' as in KKLT to get a dS solution, since it is unlikely
that such a corrected potential would have a stable critical point
and indeed that is what Choi et al find. As they have argued, the
point is that the mass of the field that is integrated out depends
on the light field and thus it cannot be integrated out as suggested
by KKLT. Our argument above highlights this point directly by showing
that the procedure of KKLT ignores effects that simply cannot be set
to zero or assumed to be small. 

Choi et al go on to analyze models with complex structure moduli.
However the analysis is done by assuming that the complex structure
moduli can be integrated out holomorphically (resulting in a holomorphic
superpotential) to get a potential in just $S$ and $T$. We will
find that this procedure is not consistent and has the same problems
that we highlighted before.

The Kaehler potential is now\begin{equation}
K=-\ln(S+\bar{S})-3\ln(T+\bar{T})+k(z^{i},\bar{z}^{\bar{j}}).\label{Kwithcs}\end{equation}
 Here $k=-\ln\int\Omega\wedge\bar{\Omega}$ (with $\Omega$ being
the holomorphic 3-form on the Calabi-Yau space) is the Kaehler potential
on the complex structure moduli space (with complex coordinates $z^{i},\, i=1,...,h_{21}$).
Also we have assumed that there is only one Kaehler structure. The
superpotential is taken to be \begin{equation}
W=A(z^{i})+SB(z^{i})+Ce^{-aT}.\label{Wwithcs}\end{equation}

The Kaehler derivatives with respect to the chiral scalars are,\begin{eqnarray}
D_{T}W & = & -aCe^{-aT}-\frac{3}{T+\bar{T}}W,\label{FS}\\
D_{S}W & = & B-\frac{W}{S+\bar{S}},\label{FT}\\
D_{i}W & = & \partial_{i}A+S\partial_{i}B+\partial_{i}kW.\label{Fz}\end{eqnarray}

Thus there are $h_{12}+2$ complex equations for as many complex variables
($h_{12}$ complex structure moduli, one Kaehler modulus and the dilaton-axion)
so that all of them can be fixed. Choi et al assume that the equation
$D_{i}W=0$ can be solved holomorphically, giving %
\footnote{Choi et al ignore the $T$ dependence but we keep it here, in any
case the relevant issue is holomorphy.%
} an effective theory for $S$ and $T$ with a superpotential\[
W=W_{eff}+Ce^{-aT}\]
where \[
W_{eff}=A(z^{i}(S,T))+B(z^{i}(S,T))S.\]

and a Kaehler potential \begin{equation}
K=-\ln(S+\bar{S})-3\ln(T+\bar{T})+k(z(S,T),\bar{z}(\bar{S},\bar{T}))\label{Keff}\end{equation}
 The SUSY conditions in the effective theory are, 

\begin{eqnarray*}
D_{S}W & = & D_{S}W|_{z^{i}}+\frac{\partial z^{i}}{\partial S}D_{i}W=0\\
F_{T} & = & D_{T}W|_{z^{i}}+\frac{\partial z^{i}}{\partial T}D_{i}W=0\end{eqnarray*}
 which are of course implied by the equations of the original theory
$D_{S}W=D_{T}W=D_{i}W=0$, with the chiral fields being all independent
variables. However this equivalence is guaranteed only if we do not
ignore the last term in (\ref{Keff}). In Choi et al however the effective
Kaehler potential is taken to be just the first two terms of (\ref{Keff}).
This is not really consistent and it is in fact the dependence of
$k$ on $z^{i}$ as well as $\bar{z^{\bar{i}}}$ that makes it impossible
to find an holomorphic solution for $S$and $T$ in terms of $z^{i}$
and hence a holomorphic $W_{eff}$. To see this consider the equation
that needs to be solved,\begin{equation}
D_{i}W=\partial_{i}A(z^{i})+S\partial_{i}B(z^{i})+W(S,T,z^{i})k_{i}=0.\label{Fzeff}\end{equation}
This is supposed to have solutions $z^{i}=z^{i}(S,T)$ such that $\partial_{\bar{S}}z^{i}=\partial_{\bar{T}}z^{i}=0$
for some range of values of $S$ and $T$. So differentiating (\ref{Fzeff})
with respect to $\bar{S}$ we get from the assumed holomorphicity,\[
W(S,T,z^{i}(S,T))k_{i,\bar{j}}\frac{\partial\bar{z}^{\bar{j}}}{\partial\bar{S}}=0.\]
 But the superpotential should not vanish (except at particular points)
and $k_{i\bar{j}}$ is the Kaehler metric on the complex structure
moduli space which is non-degenerate. Hence the above equation implies
$\frac{\partial z^{i}}{\partial S}=0$ and similarly $\frac{\partial z^{i}}{\partial T}=0$!
Clearly what is at fault is the assumption of holomorphicity. In other
words the solution of (\ref{Fzeff}) must be of the form $z^{i}=z^{i}(S,T,\bar{S},\bar{T})$.
As we saw explicitly in the case without complex structure moduli
where $S$ was integrated out, in supergravity, fields cannot be integrated
out in a holomorphic fashion. As in that case, we expect that the
effective supergravity theory is one with a superpotential that is
unity and a Kaehler potential\[
G=K=-\ln(S+\bar{S})-3\ln(T+\bar{T})+k(z(S,T,\bar{S},\bar{T}),\bar{z}(\bar{S},\bar{T},S,T))+\ln|W(S,T,z(S,T,\bar{S},\bar{T}))|^{2}.\]
 Clearly similar remarks would apply to an effective theory that is
obtained by integrating out both the complex structure moduli as well
as the dilaton-axion to get an effective theory for $T$.

\section{Effective potential for T}

Let us now try to find the effective potential for the modulus $T$
(assumed light) after integrating out the complex structure moduli
$z_{i}$ and the dilaton-axion $S$ which we assume to be heavy. Such
an effective potential is useful for cosmological applications. Hitherto
it has been derived using the two stage process of KKLT, but as we
have already seen in section 3 even in the absence of the $z_{i}$
there are terms in the effective potential that are as large as the
terms that are kept in the two stage argument but were ignored there.

The potential below the string and Kaluza-Klein scale is of the standard
$N=1$ SUGRA form with the Kaehler and superpotentials being given
respectively by eqns (\ref{Kwithcs}) (\ref{Wwithcs}). To classically
integrate out the $z_{i}$and $S$ we need to solve eqns (\ref{FS})
(\ref{Fz}) for these variables in terms of $T$ and then plug those
solutions into the expression for the potential given in (\ref{potential})
and (\ref{G}). However the equations to be solved are non-linear
in the $z_{i}$ so the best we can do is to write the general form
of the solution in a power series expansion in $Ce^{-aT}$ for $aT>>1$.
So we write,\begin{eqnarray}
S & = & \alpha+\beta Ce^{-aT}+\gamma\bar{C}e^{-a\bar{T}}+...\nonumber \\
z^{i} & = & \alpha^{i}+\beta^{i}Ce^{-aT}+\gamma^{i}\bar{C}e^{-a\bar{T}}+...\label{Szexpn}\end{eqnarray}

where the ellipses denote higher order terms in $Ce^{-aT}$. The coefficients
$\alpha,\beta,\gamma$ are functions of the (integer) fluxes. To compare
with the KKLT two stage calculation we actually need to keep terms
up to second order. If we plug this expansion into the expression
for $G$ we get

\begin{equation}
G=\ln(v+bCe^{-aT}+\bar{b}\bar{C}e^{-a\bar{T}}+cC^{2}e^{-2aT}+\bar{c}\bar{C}^{2}e^{-2a\bar{T}}+d|C|^{2}e^{-a(T+\bar{T})}+...)-3\ln(T+\bar{T})\label{Gfinal}\end{equation}
with the new constants (note that $v,\, d$ are real) being functions
of the ones in (\ref{Szexpn}) and hence of the flux integers. Calculating
the potential using (\ref{potential}) then gives\begin{equation}
V=\frac{1}{(T+\bar{T})^{2}}[a(bCe^{-aT}+2cC^{2}e^{-2aT}+c.c.)+a|C|^{2}((4\frac{a|b|^{2}}{v}-3ad)\frac{T+\bar{T}}{3}+2d)e^{-a(T+\bar{T})}]\label{finalpotential}\end{equation}

Note that the terms $cC^{2}e^{-2aT}+c.c.$ would not have been present
if we had done the calculation in two stages as in \cite{Kachru:2003aw}.
The expression is also different even in the real direction of the
potential since now we have more parameters. Since the functional
dependence of the parameters in the potential on the flux integers
is hard to evaluate explicitly and in any case is model dependent,
we believe that the only real test of the implications of the potential
coming from type IIB flux compactifications is to confront this general
form of $V$ with experiment/observation.

In fact one of the immediate consequences of the above form of the
potential is that it is possible to find metastable deSitter minima.
A simple example (with just one condensate) is illustrated in the
figures below. This example has the following parameters for (\ref{finalpotential}):
$a=\frac{2\pi}{320},\, v=0.22941751641574312,\, b=1,\, c=-1.4097828718993035,\, d=15.786002156414208,$and
$C=1$. The minimum is at $\Re T_{min}=117.138,\,\Im T=0,\, V_{min}=10^{-15}$
with $M_{p}=1$.

~~~~~~~~~~~~~~~~~~\includegraphics{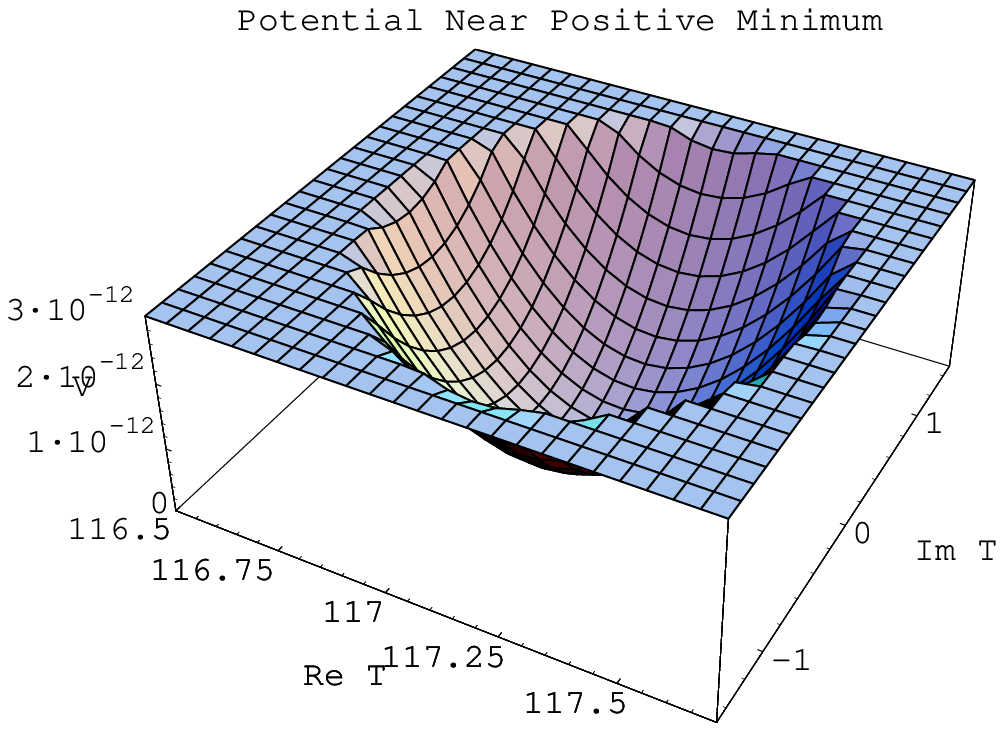}

~~~~~~~~~~~~~~~~~~~~

~~~~~~~~~~~~~~~~~~\includegraphics{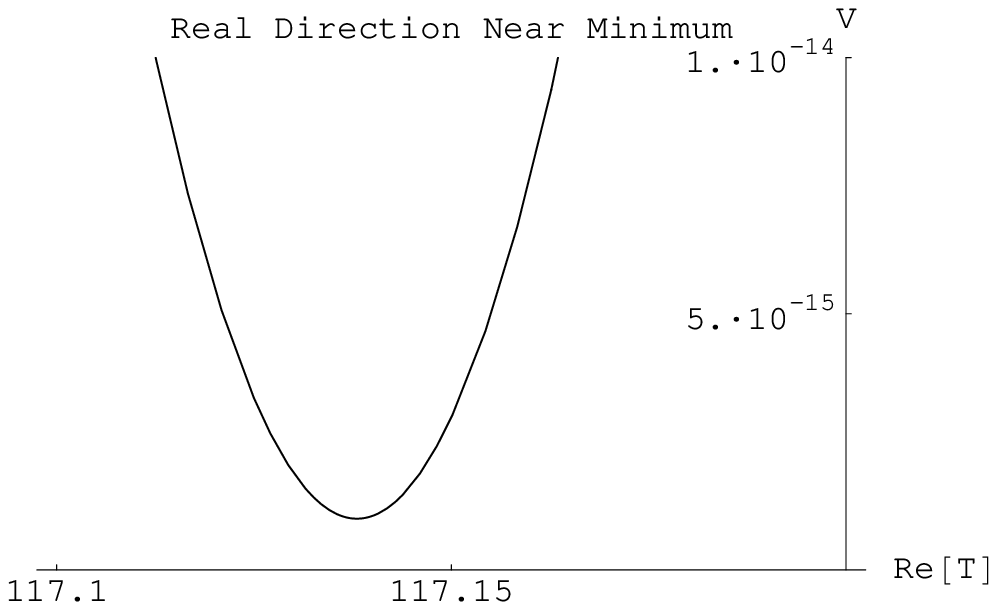}

~~~~~~~~~~~~~~~~~~~~~~~~

~~~~~~~~~~~~~~~~~~\includegraphics{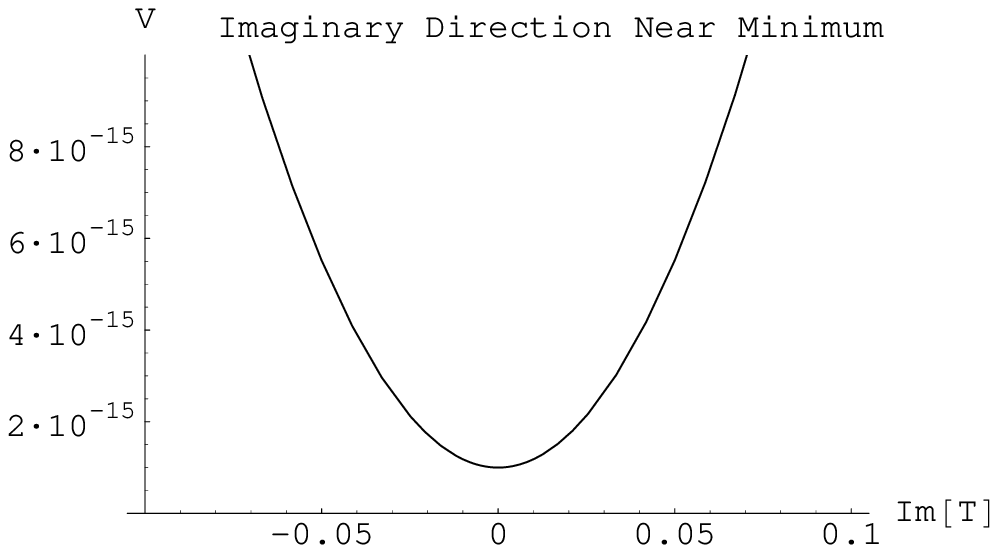}

Such potentials may also have a supersymmetric AdS minimum at $G_{T}=0$
though in this particular example this seems to be absent. Also as
is the case for all moduli potentials in string theory, for large
$\Re T$ the potential goes to zero and the positive minimum is only
meta-stable. 

A comment about the parameters chosen for our example is in order
here. The parameters are chosen to give a positive minimum at a reasonable
large value of ($\Re T\sim10^{2}$) such that also the expenential
factor $e^{-a\Re T}\sim0.1$ is small thus justifying the perturbation
expansion of adding instanton corrections to the classical superpotential.
Of course to get a critical point from such an expansion obviously
two or more of such an expansion have to be of the same order. Thus
here the term $de^{-a(T+\bar{T})}\sim0.16$ is of the same order as
constant $v\sim.2$ and $be^{-a\Re T}\sim0.1$. This doesn't necessarily
mean that the perturbation expansion is violated. It just means that
some of these coefficients need to be fine tuned in order to get a
critical point though the generic coefficients would be expected to
be of $O(1)$. This is inevitable in any such calculation of the KKLT
type (in the original calculation $W_{0}\sim Ce^{-aT}$at the critical
point and is anomalously small) since the existence of a critical
point requires that the classical terms be balanced by the non-perturbative
correction terms. For perturbation theory to be violated one would
need the coefficients of the higher order terms to continue to grow
like $e^{a\Re T}$and this is extremely unlikely.%
\footnote{This is reminiscent of a well known argument in large N guage theory
- the so-called Banks-Zachs fixed point - which is obtained by cancelling
different orders in perturbation theory, but justified on the grounds
that the resulting fixed point is still at small coupling. The argument
being that one of the coefficients of the perturbation expansion is
anomalously large thus giving the cancellation required to get a fixed
point, but that the higher order terms could still be expected to
have coefficients that did not grow with the power of the coupling
constant.%
}

So far we have worked with just one condensate. If we have several
(so that $W\sim\sum C_{i}e^{-a_{i}T}$) then one would need to make
the following replacements\begin{eqnarray*}
abCe^{-aT} & \rightarrow & \sum_{i}a_{i}b_{i}C_{i}e^{-a_{i}T}\\
acC^{2}e^{-2aT} & \rightarrow & \sum_{ij}(a_{i}+a_{j})c_{ij}C_{i}C_{j}e^{-(a_{i}+a_{j})T}\\
a^{2}|b|^{2}|C|^{2}e^{-a(T+\bar{T})} & \rightarrow & \sum_{ij}a_{i}a_{j}b_{i}\bar{b}_{j}C\bar{C}_{j}e^{-(a_{i}T+a_{j}\bar{T})}\\
a^{2}d|C|^{2}e^{-a(T+\bar{T})} & \rightarrow & \sum_{ij}a_{i}a_{j}d_{ij}C_{i}\bar{C}_{j}e^{--(a_{i}T+a_{j}\bar{T})}\\
2ad|C|^{2}e^{-a(T+\bar{T})} & \rightarrow & \sum_{ij}(a_{i}+a_{j})d_{ij}C_{i}\bar{C}_{j}e^{--(a_{i}T+a_{j}\bar{T})}\end{eqnarray*}

Finally we note that if one includes the leading perturbative correction
and non-perturbative corrections to the Kaehler potential then one
would need to replace $-3\ln(T+\bar{T})\rightarrow-3\ln(T+\bar{T}+f+ke^{-a(T+\bar{T})})$
with $f,\, k$ constants. Such an addition will clearly not change
the qualitative features of the potential.

\section{Conclusions}

In this note we examined the validity of the KKLT procedure of first
classically integrating out the dilaton-axion and the complex structure
moduli, to obtain an effective theory for the Kaehler moduli, and
then adding a non-perturbative term to the superpotential to obtain
a potential that stabilizes the Kaehler moduli. We find that there
is no approximation scheme in which the procedure of first integrating
out the $S$ and $z^{i}$ fields classically and then adding a $T$-dependent
non-perturbative term to the superpotential is justified. The latter
term needs to be included from the beginning and gives additional
contributions to the potential that cannot be ignored. Also we find
that the procedure cannot be done holomorphically, i.e. the effective
theory has to be defined entirely in terms of a Kaehler potential
(in effect the Kaehler invariant function $G$) and a superpotential
that is just unity. These considerations of course do not affect the
result of KKLT that the Kaehler modulus can be stabilized by non-perturbative
effects. But it does change the form of the potential for the Kaehler
modulus so that the physical effects (in particular the cosmological
considerations) emerging from the theory need to be reconsidered.
For instance in \cite{Brustein:2004xn} it was shown that if one follows
the KKLT procedure, then there is no way of getting a broken supersymmetric
minimum with a positive or zero cosmological constant with just one
light modulus $T$, (even with an arbitrary number of non-perturbative
terms) without adding an uplifting term. However this is no longer
the case if one correctly integrates out the heavy moduli, and we
showed in an explicit example that it is possible to obtain a positive
local minimum with just the F-term potential. Also the cosmological
considerations based on KKLT such as \cite{Kachru:2003sx}\cite{Blanco-Pillado:2004ns},
need to be revisited in light of the present results. In addition
to the effects considered here, perturbative corrections to the Kaehler
potential also need to taken into account. A complete treatment of
the physics of such models, the possibility of getting small supersymmetry
breaking, a small cosmological constant, sufficient inflation etc.
after including these corrections, will be discussed in forthcoming
work %
\footnote{R. Brustein, S.P. de Alwis and P. Martens - work in progress.%
}.

\section{Acknowledgments}

I wish to thank Ramy Brustein, Marc Grisaru, Hans-Peter Nilles and
especially Martin Rocek, for useful discussions, and Paul Martens
for the mathematica plots. This research is supported in part by the
United States Department of Energy under grant DE-FG02-91-ER-40672.

\bibliographystyle{/usr/users/dealwis/apsrev}
\bibliography{myrefs}

\begin{thebibliography}{18}
\expandafter\ifx\csname natexlab\endcsname\relax\def\natexlab#1{#1}\fi
\expandafter\ifx\csname bibnamefont\endcsname\relax
  \def\bibnamefont#1{#1}\fi
\expandafter\ifx\csname bibfnamefont\endcsname\relax
  \def\bibfnamefont#1{#1}\fi
\expandafter\ifx\csname citenamefont\endcsname\relax
  \def\citenamefont#1{#1}\fi
\expandafter\ifx\csname url\endcsname\relax
  \def\url#1{\texttt{#1}}\fi
\expandafter\ifx\csname urlprefix\endcsname\relax\def\urlprefix{URL }\fi
\providecommand{\bibinfo}[2]{#2}
\providecommand{\eprint}[2][]{\url{#2}}

\bibitem[{\citenamefont{Giddings et~al.}(2002)\citenamefont{Giddings, Kachru,
  and Polchinski}}]{Giddings:2001yu}
\bibinfo{author}{\bibfnamefont{S.~B.} \bibnamefont{Giddings}},
  \bibinfo{author}{\bibfnamefont{S.}~\bibnamefont{Kachru}}, \bibnamefont{and}
  \bibinfo{author}{\bibfnamefont{J.}~\bibnamefont{Polchinski}},
  \bibinfo{journal}{Phys. Rev.} \textbf{\bibinfo{volume}{D66}},
  \bibinfo{pages}{106006} (\bibinfo{year}{2002}), \eprint{hep-th/0105097}.

\bibitem[{\citenamefont{Kachru et~al.}(2003{\natexlab{a}})\citenamefont{Kachru,
  Kallosh, Linde, and Trivedi}}]{Kachru:2003aw}
\bibinfo{author}{\bibfnamefont{S.}~\bibnamefont{Kachru}},
  \bibinfo{author}{\bibfnamefont{R.}~\bibnamefont{Kallosh}},
  \bibinfo{author}{\bibfnamefont{A.}~\bibnamefont{Linde}}, \bibnamefont{and}
  \bibinfo{author}{\bibfnamefont{S.~P.} \bibnamefont{Trivedi}}
  (\bibinfo{year}{2003}{\natexlab{a}}), \eprint{hep-th/0301240}.

\bibitem[{\citenamefont{Kaplunovsky and Louis}(1994)}]{Kaplunovsky:1994fg}
\bibinfo{author}{\bibfnamefont{V.}~\bibnamefont{Kaplunovsky}} \bibnamefont{and}
  \bibinfo{author}{\bibfnamefont{J.}~\bibnamefont{Louis}},
  \bibinfo{journal}{Nucl. Phys.} \textbf{\bibinfo{volume}{B422}},
  \bibinfo{pages}{57} (\bibinfo{year}{1994}), \eprint{hep-th/9402005}.

\bibitem[{\citenamefont{Burgess et~al.}(1996)\citenamefont{Burgess,
  Derendinger, Quevedo, and Quiros}}]{Burgess:1995aa}
\bibinfo{author}{\bibfnamefont{C.~P.} \bibnamefont{Burgess}},
  \bibinfo{author}{\bibfnamefont{J.~P.} \bibnamefont{Derendinger}},
  \bibinfo{author}{\bibfnamefont{F.}~\bibnamefont{Quevedo}}, \bibnamefont{and}
  \bibinfo{author}{\bibfnamefont{M.}~\bibnamefont{Quiros}},
  \bibinfo{journal}{Annals Phys.} \textbf{\bibinfo{volume}{250}},
  \bibinfo{pages}{193} (\bibinfo{year}{1996}), \eprint{hep-th/9505171}.

\bibitem[{\citenamefont{Choi et~al.}(2004)\citenamefont{Choi, Falkowski,
  Nilles, Olechowski, and Pokorski}}]{Choi:2004sx}
\bibinfo{author}{\bibfnamefont{K.}~\bibnamefont{Choi}},
  \bibinfo{author}{\bibfnamefont{A.}~\bibnamefont{Falkowski}},
  \bibinfo{author}{\bibfnamefont{H.~P.} \bibnamefont{Nilles}},
  \bibinfo{author}{\bibfnamefont{M.}~\bibnamefont{Olechowski}},
  \bibnamefont{and} \bibinfo{author}{\bibfnamefont{S.}~\bibnamefont{Pokorski}},
  \bibinfo{journal}{JHEP} \textbf{\bibinfo{volume}{11}}, \bibinfo{pages}{076}
  (\bibinfo{year}{2004}), \eprint{hep-th/0411066}.

\bibitem[{\citenamefont{de~Alwis}(2005)}]{deAlwis:2005tg}
\bibinfo{author}{\bibfnamefont{S.~P.} \bibnamefont{de~Alwis}}
  (\bibinfo{year}{2005}), \eprint{hep-th/0506267}.

\bibitem[{\citenamefont{Burgess et~al.}(2003)\citenamefont{Burgess, Kallosh,
  and Quevedo}}]{Burgess:2003ic}
\bibinfo{author}{\bibfnamefont{C.~P.} \bibnamefont{Burgess}},
  \bibinfo{author}{\bibfnamefont{R.}~\bibnamefont{Kallosh}}, \bibnamefont{and}
  \bibinfo{author}{\bibfnamefont{F.}~\bibnamefont{Quevedo}},
  \bibinfo{journal}{JHEP} \textbf{\bibinfo{volume}{10}}, \bibinfo{pages}{056}
  (\bibinfo{year}{2003}), \eprint{hep-th/0309187}.

\bibitem[{\citenamefont{Brustein and de~Alwis}(2004)}]{Brustein:2004xn}
\bibinfo{author}{\bibfnamefont{R.}~\bibnamefont{Brustein}} \bibnamefont{and}
  \bibinfo{author}{\bibfnamefont{S.~P.} \bibnamefont{de~Alwis}},
  \bibinfo{journal}{Phys. Rev.} \textbf{\bibinfo{volume}{D69}},
  \bibinfo{pages}{126006} (\bibinfo{year}{2004}), \eprint{hep-th/0402088}.

\bibitem[{\citenamefont{Wess and Bagger}(1992)}]{Wess:1992cp}
\bibinfo{author}{\bibfnamefont{J.}~\bibnamefont{Wess}} \bibnamefont{and}
  \bibinfo{author}{\bibfnamefont{J.}~\bibnamefont{Bagger}},
  \bibinfo{journal}{Supersymmetry and supergravity}  (\bibinfo{year}{1992}),
  \bibinfo{note}{princeton, USA: Univ. Pr. 259 p}.

\bibitem[{\citenamefont{Gates et~al.}(1983)\citenamefont{Gates, Grisaru, Rocek,
  and Siegel}}]{Gates:1983nr}
\bibinfo{author}{\bibfnamefont{S.~J.} \bibnamefont{Gates}},
  \bibinfo{author}{\bibfnamefont{M.~T.} \bibnamefont{Grisaru}},
  \bibinfo{author}{\bibfnamefont{M.}~\bibnamefont{Rocek}}, \bibnamefont{and}
  \bibinfo{author}{\bibfnamefont{W.}~\bibnamefont{Siegel}},
  \bibinfo{journal}{Superspace, or one thousand and one lessons in
  supersymmetry - Front. Phys.} \textbf{\bibinfo{volume}{58}},
  \bibinfo{pages}{1} (\bibinfo{year}{1983}), \eprint{hep-th/0108200}.

\bibitem[{\citenamefont{Kachru et~al.}(2003{\natexlab{b}})}]{Kachru:2003sx}
\bibinfo{author}{\bibfnamefont{S.}~\bibnamefont{Kachru}} \bibnamefont{et~al.},
  \bibinfo{journal}{JCAP} \textbf{\bibinfo{volume}{0310}}, \bibinfo{pages}{013}
  (\bibinfo{year}{2003}{\natexlab{b}}), \eprint{hep-th/0308055}.

\bibitem[{\citenamefont{Blanco-Pillado et~al.}(2004)}]{Blanco-Pillado:2004ns}
\bibinfo{author}{\bibfnamefont{J.~J.} \bibnamefont{Blanco-Pillado}}
  \bibnamefont{et~al.}, \bibinfo{journal}{JHEP} \textbf{\bibinfo{volume}{11}},
  \bibinfo{pages}{063} (\bibinfo{year}{2004}), \eprint{hep-th/0406230}.

\bibitem[{\citenamefont{Balasubramanian}(2004)}]{Balasubramanian:2004wx}
\bibinfo{author}{\bibfnamefont{V.}~\bibnamefont{Balasubramanian}},
  \bibinfo{journal}{Class. Quant. Grav.} \textbf{\bibinfo{volume}{21}},
  \bibinfo{pages}{S1337} (\bibinfo{year}{2004}), \eprint{hep-th/0404075}.

\bibitem[{\citenamefont{Silverstein}(2004)}]{Silverstein:2004id}
\bibinfo{author}{\bibfnamefont{E.}~\bibnamefont{Silverstein}},
  \bibinfo{journal}{TASI / PiTP / ISS lectures on moduli and microphysics}
  (\bibinfo{year}{2004}), \eprint{hep-th/0405068}.

\bibitem[{\citenamefont{Lust et~al.}(2005)\citenamefont{Lust, Reffert,
  Schulgin, and Stieberger}}]{Lust:2005dy}
\bibinfo{author}{\bibfnamefont{D.}~\bibnamefont{Lust}},
  \bibinfo{author}{\bibfnamefont{S.}~\bibnamefont{Reffert}},
  \bibinfo{author}{\bibfnamefont{W.}~\bibnamefont{Schulgin}}, \bibnamefont{and}
  \bibinfo{author}{\bibfnamefont{S.}~\bibnamefont{Stieberger}}
  (\bibinfo{year}{2005}), \eprint{hep-th/0506090}.

\bibitem[{\citenamefont{Curio et~al.}(2005)\citenamefont{Curio, Krause, and
  Lust}}]{Curio:2005ew}
\bibinfo{author}{\bibfnamefont{G.}~\bibnamefont{Curio}},
  \bibinfo{author}{\bibfnamefont{A.}~\bibnamefont{Krause}}, \bibnamefont{and}
  \bibinfo{author}{\bibfnamefont{D.}~\bibnamefont{Lust}}
  (\bibinfo{year}{2005}), \eprint{hep-th/0502168}.

\bibitem[{\citenamefont{Choi et~al.}(2005)\citenamefont{Choi, Falkowski,
  Nilles, and Olechowski}}]{Choi:2005ge}
\bibinfo{author}{\bibfnamefont{K.}~\bibnamefont{Choi}},
  \bibinfo{author}{\bibfnamefont{A.}~\bibnamefont{Falkowski}},
  \bibinfo{author}{\bibfnamefont{H.~P.} \bibnamefont{Nilles}},
  \bibnamefont{and}
  \bibinfo{author}{\bibfnamefont{M.}~\bibnamefont{Olechowski}},
  \bibinfo{journal}{Nucl. Phys.} \textbf{\bibinfo{volume}{B718}},
  \bibinfo{pages}{113} (\bibinfo{year}{2005}), \eprint{hep-th/0503216}.

\bibitem[{\citenamefont{Balasubramanian and
  Berglund}(2004)}]{Balasubramanian:2004uy}
\bibinfo{author}{\bibfnamefont{V.}~\bibnamefont{Balasubramanian}}
  \bibnamefont{and} \bibinfo{author}{\bibfnamefont{P.}~\bibnamefont{Berglund}},
  \bibinfo{journal}{JHEP} \textbf{\bibinfo{volume}{11}}, \bibinfo{pages}{085}
  (\bibinfo{year}{2004}), \eprint{hep-th/0408054}.

\end{thebibliography}

\end{document}